\documentclass[aps,prl,twocolumn,superscriptaddress,showpacs]{revtex4}
 \usepackage{amsmath}
\usepackage{graphicx}
\usepackage{epsfig}
\newcommand{\beq}{\begin{equation}}
\newcommand{\eeq}{\end{equation}}
\newcommand{\bea}{\begin{eqnarray}}
\newcommand{\eea}{\end{eqnarray}}
\newcommand{\pdag}{{\phantom{\dagger}}}

\begin{document}
\bibliographystyle{apsrev}
 
\title{ Tunneling exponents sensitive to impurity scattering in quantum wires }
\author{  M.\ Kindermann}
\affiliation{ School of Physics, Georgia Institute of Technology, Atlanta, Georgia 30332, USA  }

\date{March 2007}
\begin{abstract}
 We show that the scaling exponent for tunneling into a quantum wire  in the  ``Coulomb Tonks gas'' regime  of impenetrable, but otherwise free, electrons is affected by impurity scattering in the wire. The  exponent for tunneling {\it into} such a wire thus depends on the conductance {\em through} the wire. This striking effect originates from a many-body scattering resonance reminiscent of  the Kondo effect.   The predicted anomalous scaling is stable against weak perturbations of the ideal Tonks gas limit at sufficiently high energies, similar to the phenomenology of a quantum critical point. 
       \end{abstract}
\maketitle

{\it Introduction:} The ``spin-incoherent'' limit of interacting quantum wires   has attracted much recent attention \cite{fiete:rmp07}. Generically this limit is reached at low electron densities, when  the Coulomb interaction suppresses spin-exchange between electrons. It appears at temperatures   higher than the spin-exchange energy $J$, when the spin configuration of the wire becomes effectively static. Many of its properties  \cite{cheianov:prl04,fiete:prl04,fiete:rmp07} differ qualitatively from those of the asymptotic  low-energy limit described by the Luttinger liquid \cite{haldane:jpc81}.
 It has been shown by Fogler that spin-incoherent physics can also be observed at high electron densities, when the Coulomb interaction  induces  the  Coulomb Tonks gas of impenetrable, but otherwise {\it free}, electrons  in ultra-thin quantum wires  \cite{fogler:prb05}.   
  
\begin{figure} 
\includegraphics[width=7.5cm]{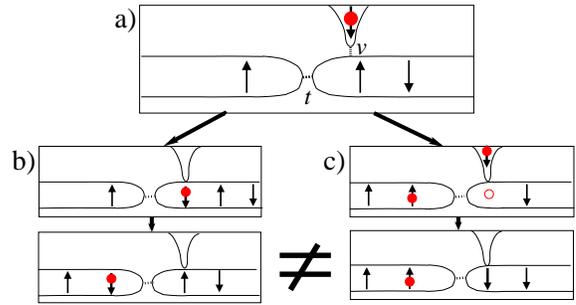}
\caption{ Tunneling (amplitude $v$) into a conductor of impenetrable electrons  near a scatterer (amplitude $t$). Panels b and c depict  processes where charge (the red balls) scatters within the wire after (b) or before (c) a tunneling event  occurs. The two processes  differ in the final spin orderings.  }   \label{fig1}
\end{figure}

In this Letter we show that, rather than being just one of many realizations of the spin-incoherent electron gas, the Coulomb Tonks gas exhibits its own and qualitatively different low-energy physics.  Like most  many-body effects observed  in quantum  wires  without spin-incoherence  \cite{bockrath:nat99,yao:nat99,auslaender:sci02,leroy:nat04} this shows in experiments where electrons tunnel  from an external probe, such as a scanning tunneling microscope tip, into the wire. We consider the situation depicted in Fig.\ \ref{fig1}, panel a, where electrons tunnel into the wire near a static impurity.  For a generic repulsively interacting electron gas, such an impurity cuts the wire into two at low energies \cite{kane:prb92}. Electrons then effectively tunnel into the end of a half-infinite conductor without any signatures of spin-incoherence \cite{fiete:rmp07}. In the Coulomb Tonks gas, in contrast, interactions are weak except for a ``hard core'' that renders the electrons impenetrable.  Electron transmission through the impurity here remains possible at low energies, ideally with an energy-independent amplitude $t$.  As a consequence,   a many-body resonance  develops that can be anticipated  already from a perturbative analysis   at weak transmission  $t\ll 1$. One finds that a perturbative calculation is invalidated at low voltages $V$ by contributions that diverge as $\ln (eV/\Lambda)$, where $\Lambda$ is the bandwidth of the  wire  (we take the zero temperature limit at even smaller $|J|$). The physics of these divergencies is analogous to that of similar divergencies in the Kondo model \cite{hewson:bo93}, and it is illustrated in panels b and c of Fig.\ \ref{fig1}: Two processes contribute to the correction of the tunneling current at lowest order in $t$. In the first, electron-like one  an electron tunnels into the wire and subsequently transmits through the impurity (Fig.\ \ref{fig1} b). In the second, hole-like  process an electron  first  scatters across the impurity and leaves a hole that   then is filled by an electron tunneling in  (Fig.\ \ref{fig1} c). The static spins  of the Coulomb Tonks gas follow the motion of charge, and thus, electrons scattering within the wire shift the spatial spin ordering. In general, the   above two processes, where this shift of the spin state occurs either before or after the addition  of a spin by the tunneling electron,  therefore result in different final spin states as illustrated in Fig \ref{fig1}.  This prevents cancellations   between the corrections to the tunneling amplitude due to these two processes   that   occur in the absence of spin. After a summation over intermediate electron and hole energies, the mentioned divergencies result.

Below we resum these divergencies in the perturbation series and find that they result in a power law divergency of the conductance  $G_{\rm tun}$  for tunneling into the wire,
 \beq \label{strongt}
 G_{\rm tun} \propto \left(\frac{\Lambda}{eV}\right)^\alpha; \;\;\; \alpha= \frac{2\arcsin |t|}{\pi}\left(1 - \frac{\arcsin |t|}{\pi}\right),
 \eeq
 where $\arcsin$ is the inverse of the $\sin$-function. The resonance described here  has the  peculiar effect of inducing  tunneling power laws  with exponents that depend on the scattering properties of and thus on the conductance through the wire. While tunneling exponents that depend on the strength of electron-electron interactions \cite{kane:prb92} or the magnetic field \cite{kindermann:prl06b} had been found before for one-dimensional quantum wires, this effect is, to the best of our knowledge, characteristic of the Coulomb Tonks gas.
   Eq.\ (\ref{strongt}) reproduces  the exponent  $\alpha=1/2$ of the impurity-free case at  $t=1$ \cite{cheianov:prl04,fiete:prl04}.

 {\it Ideal Tonks Gas:} A wire in the ideal (fermionic) Tonks gas limit is described by a one-dimensional Hubbard model with an infinite onsite repulsion $U$. This model is solved exactly in terms of a
static spin background and spinless fermions $c$, describing holes in the charge configuration \cite{ogata:prb90,kindermann:prb06}. The electron annihilation operator $\psi_\sigma$ in this solution is decomposed into the  fermions $c(x)$ and
    operators $S_{\sigma}(x)$ that add a spin 
$\sigma$ to the spin background   at position $x$ as $ \psi_{\sigma}(x)=c^\dag(x) S_{\sigma}(x)$. 
We study such a wire with an impurity at $x=0$ at zero temperature and $eV\ll\Lambda$, such that the spectrum may be linearlized,
   \bea \label{H}
  H &=& v_{\rm F}\int\frac{dq}{2\pi} \,q\, \left(c^{\dag}_{{\rm L} q} c^\pdag_{{\rm L} q} +c^{\dag}_{{\rm R} q} c^\pdag_{{\rm R} q} \right) +  H_{\rm S} ,\nonumber \\
  H_{\rm S} &=&\lambda_{\rm S}  c^{\dagger}_{{\rm L}}c^{\pdag}_{{\rm R}} + \mbox{h.c.} 
  \eea
  Here, $c_{{\rm L}q}$,  $c_{{\rm R}q}$ create electrons with momentum $q$ to the left and right of the impurity,  $c_\mu=\int (dq/2\pi) c_{\mu q}$ ($\mu \in\{\rm L,R\}$), $\lambda_{\rm S}$ is the amplitude for scattering across the impurity,  and $v_{\rm F}$  the Fermi velocity. 
  
We study  the current $I_{\rm tun}$ from a noninteracting tunnel probe  into the interacting wire at $x=0^+$ (we set $\hbar=1$),
 \bea  \label{tunnel}
   I_{\rm tun}&= &2 |v|^2 \sum_{\sigma} \int d\tau\, e^{-i e V \tau}
  \left[ {\cal G}^<_{ \sigma}(\tau){\cal G}^>_{{\rm probe},\sigma}(-\tau)\right. \\
  && \left.\mbox{}\;\;\;\;\;\;\;\;\;\;\;\;\;\;\;\;\;\;\;\;\;\;\;\;\;\;\;\;\;\;-
  {\cal G}^>_{\sigma}(\tau){\cal G}^<_{{\rm probe},\sigma}(-\tau)\right],\nonumber
\eea
where  $v$ is the amplitude for tunneling between wire and probe, and ${\cal G}_\sigma$, ${\cal G}_{{\rm probe},\sigma}$ are the Green functions of the interacting wire and the tunnel probe, respectively. We take $v\to 0$  to be arbitrarily small.  
 The Green function ${\cal G}_{\sigma}^>(\tau)=-i \langle \psi^\pdag_{\sigma}(0^+,\tau)\psi^\dag_{\sigma}(0^+,0)\rangle$   at the point of tunneling in terms of hole and spin operators takes the form
  \beq \label{Gup}
 {\cal G}_{\sigma}^>(\tau) = -i\langle S^\pdag_\sigma(0^+,\tau)c_{\rm R}^\dag(\tau)  c_{\rm R}(0)S^\dag_\sigma(0^+,0)\rangle .
\eeq
The spin expectation value in Eq.\ (\ref{Gup}) is non-vanishing only if all  spins between the one   at $x=0^+$ at time $\tau$  and the spin that is  at $x=0^+$ at time $0$ have orientation $\sigma$. This occurs with probability $p_\sigma^{|N_{\rm R}(\tau)-N_{\rm R}(0)|}$, where  $N_{\rm R}= \int(dq/2\pi) c^{\dag}_{{\rm R}q} c^{\pdag}_{{\rm R}q}  $ is the number of electrons to the right of the impurity and $p_\uparrow=1-p_\downarrow=[\exp(-
\beta E_Z)+1]^{-1}$  is the probability for a spin at inverse temperature  $\beta$ to point along  the magnetic field with Zeeman 
energy $E_Z\ll eV$. Consequently the spin and the charge expectation values in Eq.\  (\ref{Gup}) do not factorize and we obtain \cite{fiete:prl04}
 \beq \label{intG}
 {\cal G}_{\sigma}^>(\tau) =z^{-1}_{\tau} \sum_k p_\sigma^{|k|} \int_{-\pi}^{\pi} \frac{d\xi}{2\pi} e^{i\xi k}  {\cal G}_{\xi \sigma}^>(\tau),
    \eeq
   \beq \label{Gxi}
{\cal G}_{\xi\sigma}^>(\tau) = -i  \langle e^{i H \tau}  c_{\rm R}^\pdag e^{i \xi N_{\rm R} }e^{-i H \tau} e^{-i\xi N_{\rm R}}c_{\rm R}^\dag\rangle 
\eeq
after a particle-hole transformation. Here,   $z_\tau=\exp (-i E_{\rm Z} \sigma \tau/2 )$ and  ${\cal G}_{\sigma}^<(\tau) = p_\sigma z_\tau^{-2}
  {\cal G}_{\sigma}^{>}(\tau)^{*}$.

 {\it Weak Transmission:} At low bias voltages a perturbative evaluation of $I_{\rm tun}$ in $\lambda_{\rm S}$ is invalidated by the logarithmic divergencies motivated in the introduction.  
We thus start with a perturbative renormalization group (RG) approach valid at weak  $\lambda_{\rm S} \ll v_{\rm F}$, summing all contributions to the perturbation series of  leading order in the diverging logarithms at a given order in $\lambda_{\rm S}$. To compute ${\cal G}_{\xi\sigma}$, Eq.\ (\ref{Gxi}), and thus $I_{\rm tun}$ via Eq.\ (\ref{tunnel}), we add the vertices 
\beq \label{Hv}
H_{\rm v} =\sum_{k,\mu\in\{\rm L,R\}} v_{\mu k} c_{\mu}^\pdag e^{i \xi(N_{\rm R}-k) }
\eeq
to  $H$. While not directly needed for ${\cal G}_{\xi\sigma}$, the vertices with $\mu={\rm L}$ or $k\neq 0$ are included as they are generated by the RG flow. In terms of the logarithm $l=\ln (\omega/\Lambda)$ of the running cut-off $\omega$ we then find the scaling equations 
\bea \label{scaling}
dv_{{\rm R}k}/dl&=& (2\pi v_{\rm F})^{-1} \lambda_{\rm S} (v_{{\rm L}k}-v_{{\rm L}k+1}), \nonumber \\
dv_{{\rm L}k}/dl&=& (2\pi v_{\rm F})^{-1} \lambda^*_{\rm S} (v_{{\rm R}k}-v_{{\rm R}k-1}).
\eea 
Here, the first terms on the right hand sides of Eqs.\ (\ref{scaling}) derive from electron-like processes ($H_{\rm S}$  acting to the left of $H_{\rm v}$ - see Fig.\ \ref{fig1} b), and the second terms are due to  hole-like events (Fig.\ \ref{fig1} c). 
At low energies 
the eigenvector $\tilde{v}$ of the right hand sides of Eqs.\ (\ref{scaling}) with the smallest eigenvalue dominates their solution for generic initial conditions. We find  $\tilde{v}_{\rm L k}= \tilde{v}_0(-1)^k$, $\tilde{v}_{\rm R k}= \tilde{v}_0(-1)^{k+1} \lambda_{\rm S}/ |\lambda_{\rm S}|$    and the corresponding eigenvalue yields
\beq \label{scalevt}
d \tilde{v}_0/dl= - (\pi v_{\rm F})^{-1}   |\lambda_{\rm S}| \tilde{v}_0.
\eeq
 The Green function ${\cal G}_{\xi\sigma}$, Eq.\ (\ref{Gxi}), averages a product of two of the vertices contained in $H_{\rm v}$. Using the solution of Eq.\ (\ref{scalevt}) in Eqs.\ (\ref{Gxi}) and (\ref{tunnel}) we arrive, up to prefactors, at the scaling of the tunneling conductance $G_{\rm tun}=I_{\rm tun}/V$,  
  \beq \label{Iscale}
 G_{\rm tun} \sim \left(\frac{\Lambda}{eV}\right)^{2 |t|/\pi}.
 \eeq
 Here, $|t|=|\lambda_{\rm S}|/v_{\rm F}+{\cal O}(|\lambda_{\rm S}|/v_{\rm F})^2$ is the  transmission amplitude through the impurity  introduced before. For small $t$  the logarithmic divergencies encountered in a perturbation series in $t$ thus indeed  lead to the $t$-dependent tunneling exponent advertised in the introduction.
 
{\it Good Transmission:}  Eq.\ (\ref{Iscale}) is valid only at small $t\ll 1$ and thus its scaling exponent is small  and hard to measure. To access also the nonperturbative regime of good transmission $t\simeq 1$ we employ a method that was  introduced by Abanin and Levitov in Ref.\ \cite{abanin:prl04}. 
 To this end we use time translation invariance to rewrite Eq. (\ref{Gxi}) in a form directly corresponding to Eq.\ (6) of Ref.\  \cite{abanin:prl04},
 \beq \label{Gxip}
{\cal G}_{\xi\sigma}^>(\tau) = -i  \langle   c_{\rm R}^\pdag e^{-i H_\xi \tau} c_{\rm R}^\dag e^{i H \tau}\rangle , \;H_\xi=e^{i \xi N_{\rm R} }H e^{-i\xi N_{\rm R}}.
\eeq
We may thus closely follow Ref.\  \cite{abanin:prl04} in expressing Eq.\ (\ref{Gxip}) in a basis of time-dependent scattering states with  indices $s$ for their scattering time and $\gamma$ for their direction and rewriting it in terms of single-particle operators,  
 \bea \label{Green2}
{\cal G}_{\xi\sigma}^>(\tau) &=& -i\, {\rm det}\left\{[1+(T-1)f]\right\}\sum_{\gamma\gamma'} u_\gamma  u_{\gamma'}^* \\
&&  \mbox{} \times \left\{(1-f)\left[f+T^{-1}(1-f)\right]^{-1}\right\}_{0\gamma,\tau\gamma'}, \nonumber   
\eea												
where $u_\gamma $ is the amplitude for an electron in state $\gamma$ to be at $x=0^+$, $f_{ss'}=1/2\pi i(s-s'-i\delta)$ with  $\delta\approx 1/\Lambda$, and
\beq
T=e^{i h\tau} e^{i\xi n_{\rm R}} e^{-i h\tau} e^{-i \xi n_{\rm R}}.
\eeq
Here,  $h$ and $n_{\rm R}$ are the first-quantized forms of $H$ and $N_{\rm R}$, respectively. The matrix  $T$ takes the form  $T_{ss'}=\delta_{ss'} R$ for $-\tau <s<0$  and $T=\delta_{ss'}$ otherwise, where 
 \beq
  R=S^\dag e^{i\xi \sigma_R} S e^{- i\xi \sigma_R}  .
  \eeq	
  (At $\tau\geq 0$;  similarly for $\tau<0$).    $\sigma_R=(\sigma_x+i\sigma_y)/2$, with the Pauli matrices $\sigma_{x/y}$, projects onto amplitudes to the right of the scatterer that has a scattering matrix $S$.     Evaluated  in the eigenbasis of $R$, Eq.\ (\ref{Green2})   yields
  \beq \label{G0}
{\cal G}_{\xi\sigma}^>(\tau) = \frac{-i}{2\pi\delta} \sum_j |\tilde{u}_j|^2 e^{(2\chi_j/\pi-\sum_k (\chi_k/\pi)^2-1)w^2 }
\eeq
with  $ w^2=\ln[(\tau-i \delta)/(-i \delta)]$.
 The indices $j,k\in\{1,2\}$ label the two eigenvalues of $R$ and the $\tilde{u}_j$  correspond to their eigenvectors. The  phases $\chi_j$ take the form 
\beq
\chi_j=(-1)^j \arcsin\left[|t| \sin\left(\frac{\xi}{2}\right)\right].
\eeq

In the regime of weak transmission $t\ll 1$ the Green function implied by Eq.\ (\ref{G0}) has a closed form expression in terms of modified Bessel functions of the first kind $I_n$,
 \beq \label{Bessel}
{\cal G}_{\sigma}^>(\tau)= \frac{2-(2-|t|^2)\cos\varphi}{2\pi i z_\tau\delta\exp (w^2) }\sum_k (-p_\sigma)^{|k|} I_{2k}\left(\frac{2 |t| w^2}{\pi}\right).
\eeq
Here, $\varphi$ is the phase that electrons incident on the scatterer from the right pick up when backscattering.
 Without spin-polarization, when $p_\sigma\approx 1/2$,  the $k$-summation in Eq.\ (\ref{Bessel}) terminates after a few terms. Using the asymptotic form of the Bessel functions $I_n(x)\sim \exp(x)/\sqrt{2\pi x}$ at  $|t| w^2\gg \pi k^2$, that is at low  voltages, the sum over $k$ may be performed explicitly and one finds the same scaling  as with our RG approach, Eq.\ (\ref{Iscale}). 

\begin{figure} 
\includegraphics[width=7.cm]{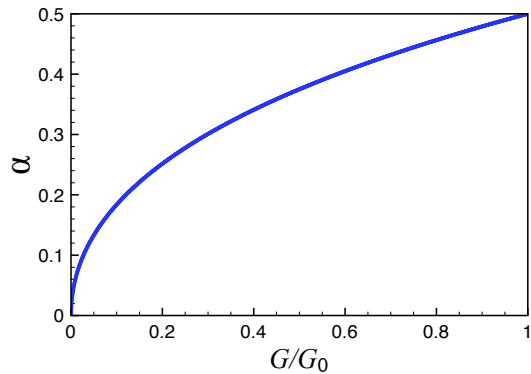}
\caption{   Scaling exponent $\alpha$ of the tunneling current {\it into} a Coulomb Tonks gas wire with impurity as a function of  the conductance $G$  {\it through} the wire, normalized to  $G_0=e^2/h$.}   \label{fig2}
\end{figure}

For general $t$ we carry out the $\xi$-integration of Eq.\ (\ref{intG})  in saddle point approximation,  again valid  at low voltages, $|t| w^2\gg \pi  $. In the absence of  spin-polarization, when $1-p_\sigma$ is not small, the dominant saddle point is located at $\xi=(-1)^j\pi$, where $j$ is the summation index in Eq.\ (\ref{G0}). The corresponding contribution to the integral is
\beq \label{mainsaddle}
{\cal G}_{\sigma}^{(1)>}(\tau)=  \frac{(1-\cos\varphi \cos \theta )\sqrt{\cot  \theta }}{  4 \pi i z_\tau w\delta \sqrt{1/2- \theta/\pi  }}\frac{ 1-p_\sigma}{ 1+p_\sigma} \left(\frac{-i \delta}{\tau-i \delta}\right)^{1-\alpha } ,
\eeq
 with  $\alpha=2 (1 -\theta/\pi )\theta/\pi$  and $\theta = \arcsin |t|$. Using   Eq.\ (\ref{tunnel}) one immediately arrives at our main result, Eq.\ (\ref{strongt}). Experimentally $|t|$ can be inferred from a measurement of the conductance through the interacting wire $G=|t|^2 e^2/h$ if the scatterer at $x=0$ is the only impurity in the wire. For that case we show the dependence of the exponent $\alpha$ on $G$  over its entire range in Fig.\ \ref{fig2}. 
 The cross-over to spin-polarized behavior in a magnetic field is described by a second saddle point of the $\xi$-integral in Eq.\ (\ref{intG}),
 \beq
 {\cal G}_{\sigma}^{(2)>}(\tau)=\frac{1}{\sqrt{2\pi}\pi z_\tau} (1-\cos \varphi \cos \theta)  \frac{1}{\tau-i \delta},
 \eeq
 valid at $1-p_\sigma \ll 1$. The crossover of the scaling of $G_{\rm tun}$ to Ohmic behavior takes place as the Zeeman energy exceeds the crossover scale
 \beq
 E_Z^* =  -2 \beta^{-1} \frac{\theta}{\pi}\left(1-\frac{\theta}{\pi}\right) \ln \left(\frac{\Lambda}{eV}\right).
 \eeq
 The crossover occurs at decreasing magnetic fields as the transmission through the wire decreases until at $t=0$ the tunneling current is Ohmic already in zero field. 
 
 {\it Coulomb Tonks Gas:} In the Coulomb Tonks gas formed by real electrons in quantum wires, the tails of the Coulomb potential introduce weak forward scattering interactions in addition to the hard core repulsion that renders the electrons impenetrable \cite{fogler:prb05}.   The dynamics of the charge carrying holes in the interacting wire may then be described by a Luttinger liquid with interaction parameter $g$.  The ideal Tonks gas considered above is recovered at $g=1$. At sufficiently high energies the forward scattering interactions are weak also on the Coulomb Tonks gas, $|1-g| \ll 1$.    Hence we may include these interactions in our earlier perturbative RG approach valid at $t\ll 1$. We  first take $g$ to be energy-independent. The flow equations (\ref{scaling}) in that case have to be complemented by an additive contribution $(g^{-1}-1)v_{\mu k}/2$ to $dv_{\mu k}/dl$ and an additional equation $dt/dl=(g^{-1}-1)t$, yielding
 \beq \label{intscale}
  G_{\rm tun} \sim \left(\frac{eV}{\Lambda}\right)^{1-g} \exp\left[\frac{2 |t|}{\pi}\,\frac{1-(eV/\Lambda)^{1-g}}{1-g}\right]
 \eeq
 at $|1-g|\ll 1$ (up to a prefactor). Expanding the power law inside the exponential of Eq.\ (\ref{intscale}), we find that the scaling of Eq.\ (\ref{Iscale}) is still observable over $\lesssim |1-g|^{-1}$ decades, with a renormalized transmission $t_V$ and an additional renormalization of the tunneling amplitude $v$,
\beq \label{ratio}
  \frac{G_{\rm tun}(V)}{ G_{\rm tun}(V_0)} \sim \left(\frac{V}{V_0}\right)^{1-g-2 |t_{V}|/\pi }, \;\;t_V=\left(\frac{eV}{\Lambda}\right)^{1-g} t,
 \eeq
for  $|\ln(V/V_0)| \ll |1-g|^{-1}$.
 We expect that similarly  also Eq.\ (\ref{strongt}) at strong transmission is robust against the weak forward scattering interactions in the Coulomb Tonks gas   with a restricted  scaling range of voltages. At $V=0$, $  G_{\rm tun}/G_{\rm tun}|_{t=0}$ has an essential singularity at $g=1$  [see Eq.\ (\ref{intscale})]. This  formally  supports our earlier statement that the ideal  Tonks gas limit  qualitatively differs from  the general spin-incoherent electron gas with $g\neq 1$. The above nonanalyticity at $V=0$ is an indication of possible nonanalyticities in ground state properties and thus of a quantum phase transition in our model at $g=1$.  This conjecture finds further support in the scaling behavior of $G_{\rm tun}$. Perfect scaling  has been found only at the (possibly critical) point $g=1$. But for sufficiently high energies,   $eV\gg eV_g=\Lambda\exp(-1/|1-g|)$, scaling persists to $g\neq 1$,  the typical behavior around a quantum critical point \cite{sachdev:bo99}. One may thus speculate that there is a phase transition in our model between two qualitatively different ground states, where i)  all spins in the interacting wire are entangled with the noninteracting probe ($g>1$) and ii)  this holds only for one half of the wire that is cut into two by the impurity ($g<1$). Definite statements about the ground state, however, are not possible on the basis of the above calculation, because it is perturbative in the amplitude $v$ that grows large in the limit $V\to 0$.  

 The Coulomb Tonks gas is realized in wires whose  radius $R$  is much smaller than their effective Bohr radius $a_B=\hbar^2 \kappa/m_* e^2$, such that the parameter ${\cal L}=\ln(a_B/R)$ grows large  \cite{fogler:prb05}.  Here,  $m_*$  is the effective mass of the conduction electrons, and $\kappa$ is the dielectric constant of the substrate that supports the wire. The regime  appears at high electron density $n$, when   $r_{\rm s}=1/2na_B$ is small, ${\cal L}^{-1}\ll r_{\rm s}\ll 1$. The electrons in such a wire are impenetrable, with $J\approx \Lambda [r_{\rm s}({\cal L}+\ln r_{\rm s})]^{-1}$, but otherwise free at energies $\omega \gg \omega^*$ exceeding an exponentially small  scale $  \omega_* =(\Lambda/r_{\rm s}) \exp(-\pi^2/2r_{\rm s})$.  The forward scattering part of the Coulomb potential is indeed a small perturbation described by an interaction parameter $1-g_{\rm C}(\omega) \approx (2r_{\rm s}/\pi^2) \ln (\Lambda/\omega) \ll 1$ \cite{fogler:prb05}. Repeating our above RG-analysis for this energy-dependent interaction, we find $v_{\mu k} \sim \exp[-r_{\rm s} l^2/2 \pi ^2- |t| \sqrt{\pi/4 r_{\rm s}} \text{erf}\left(l  \sqrt{r_{\rm s}}/\pi\right) ] $.  For frequency ratios $\ln(\omega/\omega_0)\ll {\rm min} \{[1-g_{\rm C}(\omega)]^{-1},\pi/\sqrt{r_{\rm s}}\}$ we may expand   the error function ${\rm erf}$ around $l=\ln(\omega_0/\Lambda)$ to find for $G_{\rm tun} $ an expression like Eq.\ (\ref{ratio}) with  $g=g_{\rm C}(eV)$ and  $t_V=(eV/\Lambda)^{(1-g)/2}t$.  At $t \ll 1$ the predicted scaling is thus observable over  $ \lesssim{\rm min} \{\pi^2/2r_{\rm s} \ln (\Lambda/eV),\pi/\sqrt{r_{\rm s}}\}$ decades with $eV\gg \beta^{-1}$. For the above effects to be observable the  spin-incoherent condition $\beta^{-1}\gg J$ has to be satisfied. If tunneling takes place remote from the scatterer, at $x=x_0\neq 0$,  $eV\ll v_F/ x_0$ is required additionally. Our calculation further assumes $eV \ll \Lambda_0$, where  $\Lambda_0$ is the width of the energy window around the Fermi level in which the energy-dependence of the scattering amplitude $\lambda_{\rm S}$   may be neglected. The above scaling    thus occurs at $  \Lambda /r_{\rm s}({\cal L}+\ln r_{\rm s}) \ll \beta^{-1}\ll eV \ll   {\rm min} \{v_F/x_0,\Lambda_0\}$.  Alternatively, the Coulomb Tonks gas is found in gated wires at $R\lesssim a_{\rm B}$ and low densities, $n \ll a_{\rm B}/D^2 \ll a^{-1}_{\rm B}$, where $D$ is the distance between wire and gate \cite{haeusler:prb02,fogler:prb05}. 
 
 The physics described above is relevant also  for ultracold clouds of fermionic atoms. There the Tonks gas regime $g\approx 1$ is naturally realized since atoms are charge neutral and interact only via a local, ``hardcore'' potential. The bosonic Tonks gas  \cite{paredes:nat04,kinoshita:sci04} as well as one-dimensional Fermi gases  \cite{moritz:prl05} have already been demonstrated experimentally. Progress is being made in cooling Fermi gases substantially below the Fermi temperature \cite{demarco:sci99,truscott:sci01}, as required to observe the scaling predicted above. Also the experimental techniques for transport measurements with local probes as described above have been proposed \cite{thywissen:prl99,micheli:prl04} and are expected to be experimentally implemented in the near future. 
 
 {\it Conclusions:} We have predicted a many-body scattering resonance for the strongly correlated Tonks gas regime of quantum wires. It is reminiscent of the Kondo resonance and possibly marks the  critical point of a quantum phase transition. The effect occurs in the presence of static impurities and it has  a clear and intriguing  experimental signature: The tunneling current {\em into}  the wire  obeys a power law with an exponent that depends on the  conductance {\em through} the wire. This anomalous scaling is observable in ultra-thin wires at high electron density, in gated conductors at low density, and in ultracold gases of fermionic atoms.
 
 The author thanks M.\ M.\ Fogler very much for helpful correspondence.

  \end{document}